\documentclass[11pt,preprint,superscriptaddress,nofootinbib]{revtex4}

\usepackage{psfig}
\usepackage{epsfig}
\usepackage{color}

\def\beq{\begin{equation}}
\def\eeq{\end{equation}}
\def\beqar{\begin{eqnarray}}
\def\eeqar{\end{eqnarray}}

\def\pfrac#1#2{\left( \frac{#1}{#2} \right)}

\def\iso#1#2{\mbox{${}^{#2}{\rm #1}$}}
\def\k4#1{\iso{K}{4#1}}
\def\u23#1{\iso{U}{23#1}}
\def\th23#1{\iso{Th}{23#1}}

\def\dens{\tilde{\rho}}
\def\nuebar{\mbox{$\bar{\nu}_e$}}

\def\la{\mathrel{\mathpalette\fun <}}
\def\ga{\mathrel{\mathpalette\fun >}}
\def\fun#1#2{\lower3.6pt\vbox{\baselineskip0pt\lineskip.9pt
  \ialign{$\mathsurround=0pt#1\hfil##\hfil$\crcr#2\crcr\sim\crcr}}}

\begin{document}

\title{Imaging the Earth's Interior:
the Angular Distribution of Terrestrial Neutrinos}

\author{Brian D. Fields\footnote[3]{{\tt bdfields@uiuc.edu}}}
\affiliation{Center for Theoretical Astrophysics,
Department of Astronomy, University of Illinois}
\affiliation{Department of Physics, University of Illinois}

\author{Kathrin A. Hochmuth\footnote[1]{{\tt hochmuth@uiuc.edu}}}
\affiliation{Department of Physics, University of Illinois}

\begin{abstract}
Decays of radionuclides throughout the Earth's interior
produce geothermal heat, but also are a source of antineutrinos;
these geoneutrinos are now becoming observable
in experiments such as KamLAND.  
Heretofore, the (angle-integrated) geoneutrino flux has
been shown to provide a unique probe of geothermal heating due to
decays, and an integral constraint on the distribution of
radionuclides in the Earth. In this paper, we calculate the angular
distribution of geoneutrinos, which opens a window on the {\em
differential} radial distribution of terrestrial radionuclides.  We
develop the general formalism for the neutrino angular distribution.
We also present the inverse transformation which
recovers the terrestrial radioisotope distribution
given a measurement of the neutrino angular distribution.
Thus, geoneutrinos not only allow a means to image the
Earth's  interior, but
offering a direct measure of the radioactive Earth,
both (1) revealing the Earth's inner structure as probed
by radionuclides, and (2) allowing for a complete determination of the
radioactive heat generation as a function of radius.
Turning to specific models, we
emphasize the very useful approximation in which the Earth is
modeled as a series of shells of uniform density. Using this
multishell approximation, we present the geoneutrino angular distribution for
the favored Earth model which has been used to calculate geoneutrino
flux.  In this model the neutrino generation is dominated by decays of
potassium, uranium, and thorium in the Earth's mantle and crust;
this leads to a very
``peripheral'' angular distribution, in which 2/3 of the neutrinos
come from angles $\theta \ga 60^\circ$ away from the downward vertical.
We note that a measurement of the neutrino intensity in peripheral directions
leads to a strong lower limit to the
central intensity.  We also note that there is some controversy about
the abundance of potassium in the Earth's core; different geophysical
predictions lead to strongly varying, and hence distinguishable, central
intensities ($\theta \la 30^\circ$ from the downward vertical).  
Other
uncertainties in the models, and prospects for observation of the
geoneutrino angular distribution, are briefly discussed.
We conclude by urging the development and construction
of antineutrino experiments
with angular sensitivity.
\end{abstract}

\maketitle

\section{Introduction}

The decays of radioactive species within the Earth
generate an important component of geothermal heat.
However, a quantitative accounting of the radioactive
energy generation of the Earth
requires a detailed knowledge of the abundance
distribution of the key long-lived radioisotopes---uranium,
thorium and potassium--inside the Earth.
Because our knowledge of these abundance distributions
is incomplete, the radiogenic heat output is consequently
model-dependent and thus uncertain.
A fundamental diagnostic is the ``Urey ratio'' 
which measures the ratio $P_{\rm rad}/P_{\rm lost}$
of total radioactive heat production to the surface heat loss. Current estimates span the range
{\cite{stein}, \cite{richter}
\beq
\label{eq:urey}
P_{\rm rad}/P_{\rm lost} \sim 0.5-0.6
\eeq
which seems to suggest that radiogenic heating is a dominant
heat source, but not the only one.
It is even possible that the Urey ratio is closer to 1,
as the primordial
heat from Earth's formation should have radiated away a long time ago
\cite{france}, and so the Earth's heat should be fully
radiogenic.  
This would imply that more radioactive material is hidden in the Earth.
The strength of these qualitative conclusions thus hangs
on the strength of the quantitative measurements of the radiogenic
heat production and total heat loss.

A beautiful if challenging
means of measuring the radiogenic heat production of
the Earth follows from the realization that
$\beta$-decays not only are a heat source but also
produce (anti)-neutrinos.
Decades ago,
the prescient work of
Eder \cite{eder} and later Krauss, Glashow, and Schramm \cite{kgs}
pointed out
that the radioactive heat flux and
the neutrino flux from the Earth are tightly linked.
A measurement of the neutrino flux would
constrain the radiogenic heat production,
and thus offer a new and direct measure of the Urey ratio.
Clearly, a precise measurement of the Urey ratio
would provide important insight into the
interior structure and dynamics of the Earth.

This vision of neutrino geophysics has enjoyed a major advance
recently, with the first detection of geoneutrinos
by the KamLAND Collaboration
\cite{kamland}. In fact, the detection of these geoneutrinos was only a side
product of the KamLAND detector, as it is primarily designed to detect
the flavor change of reactor
antineutrinos produced by the Japanese nuclear power plants. The
KamLAND Collaboration reported a total of 9 geoneutrino events,
of which they estimate 4 are from \u238 decay, and 5 are from
\th232 decay \cite{kamland}.
Obviously these are early days, but we are encouraged
that, even in the prescence of a dominant anthropogenic background,
KamLAND has demonstrated that geoneutrinos exist
at observable levels.

Several groups have taken interest in these neutrinos and have in
conjunction with geophysicists and geologists worked on models for the distribution of radioactive elements in the Earth and predicted the
(angle-integrated) geoneutrino
flux at different detector sites \cite{rcc,fmr,flmrs,mcfl,ntf}.
The flux calculations have already become rather sophisticated,
and are based on detailed geophysical models,
in some cases even including the effects of the
anisotropic radioisotope densities in the crust.
These models confirm in detail that indeed
the geoneutrino heat flux is proportional to radiogenic heat flux,
but with the important caveat that
the exact scaling between the two
depends on the detailed abundance and density
distributions within the Earth.\footnote{
It is true that, at least in a spherically symmetric Earth,
the surface heat flux is directly related, via Gauss' law,
to the total mass of radionuclides (c.f.~eq.~\ref{eq:heat}).
However, a simple Gauss' law argument
fails in the case of neutrinos, because at each point they
are emitted isotropically, not just radially,
and as we will see, the ``sideways'' neutrinos make a large
contribution to the total surface geoneutrino flux.}
In this sense, the geoneutrino flux measurement is
an {\em integral} measure of the radioisotope distribution.

In this paper we want to consider the angular distribution of the
geoneutrino flux.
We show that, once well-measured, the angular distribution
can be inverted to recover the full density distribution of
radionuclides--a tomography of the structure and
radiogenic heat generation of the Earth.
In addition, we come to the conclusion that with a future
low-energy antineutrino detector with even crude
angular resolution, it will be
possible to distinguish between the different Earth Models and solve
the problem as to how much radioactive material is contained in the
Earth and where it is located.
Thus, the angular distribution provides a {\em differential} measurement
of the radioisotope distribution, and can reveal
a wealth of new information about the structure and content of
the Earth.

We first present the formal calculations in \S \ref{sect:formal}.
After outlining the general formalism, we
consider the useful approximation of a uniform density shell,
from which a multiple-shell model of the Earth can be constructed.
We then review
(\S \ref{sect:earth}) models for the radioisotopic content of the
Earth. Using these models we construct physically motivated
illustrations of plausible geoneutrino angular distributions
(\S \ref{sect:results}).
We present several plots for different
geological abundance predictions, finding in general
that the high radioistope content of the crust
leads to a large ``peripheral'' geoneutrino signal.
We investigate the
possibility of \k40 in the core, as recently suggested
by several groups
\cite{rama,berc,1200}; if core \k40 is at the high end of these
predictions, then the resulting neutrino signal could be quite
large and should lead to a readily observable central intensity peak.
Finally, we present conclusions \S \ref{sect:conclude},
and discuss the exciting possibilities that will
arise when we are able to use neutrinos to image the
interior of the Earth.

Before we begin the formal development, a word of clarification
seems in order.
Note that for brevity, we will refer to the emitted particles
as neutrinos, although they are of course antineutrinos.
In fact, there is some regular $\nu_e$ production
due to \k40 electron captures.
However, the branching here is 10.72 \% of all \k40 decays,
and thus the majority of the \k40 nuclei
$\beta$-decay and yield \nuebar\ with a continuous energy spectrum.
We look forward to the day when the monoenergetic
\k40 electron capture $\nu_e$ flux (and angular distribution!)
can be measured and compared against the \nuebar\ signal.
However, in this paper we will consider only the dominant,
$\beta$-decay \nuebar\ signals from K, Th, and U.

\section{Formalism}
\label{sect:formal}

The fundamental quantity we wish to calculate is the
differential intensity $I$, or surface brightness,
of geoneutrinos at the surface of the Earth.  
This is just the distribution
of neutrino flux $F$ versus solid angle:
$I(\theta,\phi) = dF/d\Omega$.
Both angular coordinates
are local and observer-centered:
$\theta \in [0,\pi/2]$ is the nadir angle, i.e.,
the angle measured from the downward vertical
(so that the center of the Earth is at $\theta = 0$,
and the horizontal is at $\theta = \pi/2$; see Figure \ref{fig:geom}).
The angle $\phi$ is an azimuth. These
angles thus cover the ``sky'' underfoot (or rather, the terrestrial
hemisphere) which we wish to image.  In this paper
we will consider only the case of a spherically symmetric
Earth (including the outermost layers).  This guarantees
that the intensity $I = I(\theta)$ is azimuthally
symmetric and so only depends on the nadir angle.
In this case, we have
$I(\theta) = (2\pi)^{-1} dF/d\cos\theta$.

\begin{figure}[htb]
\begin{center}
\begin{picture}(0,0)%
\includegraphics{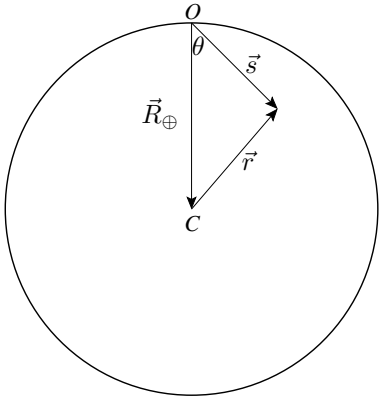}%
\end{picture}%
\setlength{\unitlength}{1184sp}%
\begingroup\makeatletter\ifx\SetFigFont\undefined%
\gdef\SetFigFont#1#2#3#4#5{%
  \reset@font\fontsize{#1}{#2pt}%
  \fontfamily{#3}\fontseries{#4}\fontshape{#5}%
  \selectfont}%
\fi\endgroup%
\begin{picture}(7846,8227)(878,-7883)
\put(4801,-736){\makebox(0,0)[lb]{\smash{{\SetFigFont{10}{12.0}{\familydefault}{\mddefault}{\updefault}{\color[rgb]{0,0,0}$\rm \theta$}%
}}}}
\put(5926,-1111){\makebox(0,0)[lb]{\smash{{\SetFigFont{10}{12.0}{\familydefault}{\mddefault}{\updefault}{\color[rgb]{0,0,0}$\vec{s}$}%
}}}}
\put(3751,-2161){\makebox(0,0)[lb]{\smash{{\SetFigFont{10}{12.0}{\familydefault}{\mddefault}{\updefault}{\color[rgb]{0,0,0}$\vec{R}_\oplus$}%
}}}}
\put(5851,-3136){\makebox(0,0)[lb]{\smash{{\SetFigFont{10}{12.0}{\familydefault}{\mddefault}{\updefault}{\color[rgb]{0,0,0}$\vec{r}$}%
}}}}
\end{picture}%

\caption{The basic geometry of the problem.
An observer at point $O$ measures
an intensity, which
sums the emission along a line of sight $\vec{s}$.
The nadir $\vec{R}_\oplus$ points to the center $C$ of the Earth,
which makes an angle $\theta$ with the
line of sight.
A given point along $\vec{s}$ is at
geocentric distance $\vec{r}$.
}
\label{fig:geom}
\end{center}

\end{figure}

The neutrino intensity in any direction depends on the
distribution of sources along that line of sight.
The governing equation is that of
radiation transfer for
neutrinos, which
is formally identical to the usual
expression for photons \cite{chandra},
although of
course the microsphysics is quite different
(the Earth is optically thin to neutrinos, but
neutrinos do undergo oscillations).
Then over a line of sight $\vec{s}$,
ignoring scattering and absorption,
the intensity changes according to
\beq
\label{eq:radtransf}
dI/ds = q(\vec{s})/4\pi
\eeq
where $q$
is the source function
at point $\vec{s}$,
which measures the local neutrino production rate per unit
volume.

For each radioisotope species $i$, this takes the form
\beq
q_i = \frac{n_i}{\tau_i} = \frac{\rho_i}{\tau_i m_i}  \ \  .
\eeq
Here $n_i$ and $\rho_i$ are the local number and
mass densities, respectively, of species $i$ in the Earth.
The  $\tau_i$ is the mean lifetime and $m_i$ is the mass
of an $i$ nucleus.
The total source is just
a superposition of all species:
$q = \sum_i q_i$.
The effects of neutrino oscillations are not yet included;
we will address this below (\S \ref{sect:realmod}).

Integrating eq.~(\ref{eq:radtransf}) over a line
of sight at nadir angle $\theta$,
we find the intensity
\beq
\label{eq:observocentric}
I(\theta) = \int_{0}^{2R_\oplus \cos \theta} q(\vec{s}) \, ds
\eeq
where $\vec{s}$ is centered on the observer,
as seen in Figure \ref{fig:geom}.

Since models of the Earth's structure and composition
are expressed in terms of the radius $\vec{r}$,
it is very useful to rewrite eq.~(\ref{eq:observocentric})
in these geocentric coordinates.
From Figure \ref{fig:geom} we see that the
$\vec{s}=\vec{r}+\stackrel{\rightarrow}{R_\oplus}$,
so that
$\vec{r}=\vec{s}-\stackrel{\rightarrow}{R_\oplus}$
and thus
\beq
s(r,\theta)=R_\oplus \cos \theta  \pm  \sqrt{r^2-(1-\cos^2\theta)R_\oplus^2}
\eeq
where $\pm$ corresponds to the far side (near side)
of the midpoint $s = R_\oplus \cos \theta$ of the line of sight.
Therefore, for a fixed
nadir angle $\theta$,
we can transform the integral to the geocentric coordinate system:
\beq
\label{eq:master}
I_i(\theta)
 = 2\int_{R_\oplus\sin\theta}^{R_\oplus} dr
     \frac{q_i(r) \, r}{\sqrt{r^2-(1-\cos^2\theta)R_\oplus^2}}
\eeq
where $q_i(r)$ can in a non-spherically symmetric case also be a
function of (geocentric) latitude and longitude.  In this case, the
factor of 2 is replaced by the sum of integrals for the
near-side and the far-side of the midpoint $s = R \cos \theta$.
In our special case of spherical symmetry,
the contribution from the near half of the sphere is the same
as the contribution from the far side.  

Equation (\ref{eq:master}) explicitly demonstrates
that the intensity $I(\theta)$ is an integral
transformation of the source distribution $q(r)$.
In fact, this mapping is a form of the
Abel transform \cite{bracewell}, which is used in deprojection
problems in both astrophysics \cite{bt}
and geophysics \cite{dahlen}.
Thus it is clear that a determination of
the intensity distribution offers a measure of
the source distribution.  
Namely, one can invert the transformation
(deproject the image) to fully recover the
complete source
distribution $q(r)$; this inversion procedure
is explicitly presented in Appendix \ref{sect:tomography}.
In other words, {\em a measurement of the angular distribution
of geoneutrinos not only yields an image of the
Earth's radioactive interior, but this image
can also be inverted to give a tomography of
the terrestrial radioisotope distribution}.
Clearly, the geoneutrino angular distribution offers
a unique and powerful probe of the interior of
the Earth.  This power will be further illustrated below,
where we will see that even a partial (low-resolution)
determination of $I(\theta)$ offers important
geophysical information.

In evaluating eq.~(\ref{eq:master}),
it will be convenient to introduce dimensionless scaled variables:
a radial fraction $x = r/R_\oplus \in [0,1]$,
a local mass fraction $a_i=\rho_i/\rho$, and
a local density measure $\dens = \rho/\bar{\rho}$ (with
$\bar{\rho} = 3 M/4\pi R_\oplus^3$ the mean Earth density).
It will also be useful to denote the nadir angle
cosine as $\mu=\cos \theta$.
We then rewrite eq.~(\ref{eq:master}) as a product of two
terms
\beq
I_i(\theta) = I_{i,0} \ g_i(\mu)
\eeq
Here the dimensionful overall magnitude is set by
\beq
\label{eq:Iscale}
I_{i,0} = 2\frac{N_i \bar{a_i} \bar\rho R_\oplus}{4\pi A_i \tau_i m_u}
\eeq
where $N_i$ is neutrino multiplicity, i.e.,
the number of geoneutrinos released per decay chain.
Values of $I_{i,0}$ appear in Table~\ref{tab:nukeprop}.
The dimensionless angular distribution or
(akin to the ``phase function'' of radiation
transfer \cite{chandra})
is the heart of this paper, and is contained in the function
\beq
g_i(\mu)=\int_{\sqrt{1-\mu^2}}^{1} dx \
   \frac{\dens_i(x) \ x}{\sqrt{x^2-(1-\mu^2)}}
\eeq
which requires knowledge of the density distribution
of each radioisotope, usually presented
in terms of mass fractions $a_i$ and a total
density profile $\rho$ via 
$\dens_i = \rho_i/\bar{\rho_i} = (a_i/\bar{a_i}) \dens$.

The flux of geoneutrinos integrated over different annuli
is also of interest.
We quantify this in terms of the flux {\em exterior} to the nadir angle
$\theta$, via
\beqar
\label{eq:flux}
F_i (> \theta) = F_i(<\mu) 
 & = & \int_{\Omega_{\rm shell}} d\Omega \ I_i(\mu,\phi) \\
 & = & 2{\pi} I_{i,0} \int_0^\mu d\mu \ g(\mu)\\
 & \equiv & 2\pi I_{i,0} H(\mu)
\eeqar
where second and third expressions assume spherical symmetry,
and where the dimensionless quantity
$H$ encodes the angular dependence.
A consequence of this definition is that
the total neutrino flux is given by
$F_i(\theta >0) = F_i(\mu < 1) = 2\pi I_{i,0} H(1)$.
This
can be compared with existing calculations
\cite{kgs,fmr,mcfl}.

\begin{table}[htb]
\caption{Properties of the Principle Geoneutrino Source Nuclei}
\label{tab:nukeprop}
\begin{tabular}{c|c|c|c|c|c}
Radioisotope & mean life & \nuebar\ multiplicity 
  & isotopic & mean terrestrial 
  & intensity normalization \\
species & $\tau$ (Gyr)  & $N_i$
  &  abundance  & abundance $\bar{a_i}$ 
  & $I_{i,0} \ [{\rm neutrinos \ cm^{-2} \ s^{-1} \ sr^{-1}}]$ \\
\hline
\k40  & 1.84 & 1 & 0.0117\% & $1.8 \times 10^{-8}$ & $2.4 \times 10^6$  \\
\u238 & 6.45 & 6 & 99.2745\% & $5.3 \times 10^{-8}$ & $0.48  \times 10^6$ \\
\th232 & 20.3 & 4 & 100\% & $1.35 \times 10^{-8}$ & $0.56 \times 10^6$ \\
\end{tabular}
\end{table}

\subsection{The Uniform Shell Approximation}

We want to consider a single shell with a density $\dens_{i,0}$
in species $i$ which is constant
between
$r_{\rm in} = x_{\rm in} R_\oplus$ and $r_{\rm out} = x_{\rm out} R_\oplus$,
and zero otherwise:
\beq
\dens_i(x) =
  \left\{
  \begin{array}{cc}
    \dens_{i,0} & x_{\rm in} < x < x_{\rm out} \ , \\
    0 \ , & \mbox{otherwise}
  \end{array}
  \right.
\eeq
This form
allows us to simplify the integral and solve it analytically.
In particular, owing to the constant density,
the intensity at each line of sight is
just proportional to the shell path length
$\Delta s(\theta)$ along that sightline.

\begin{figure}[htb]
\begin{center}
\begin{picture}(0,0)%
\includegraphics{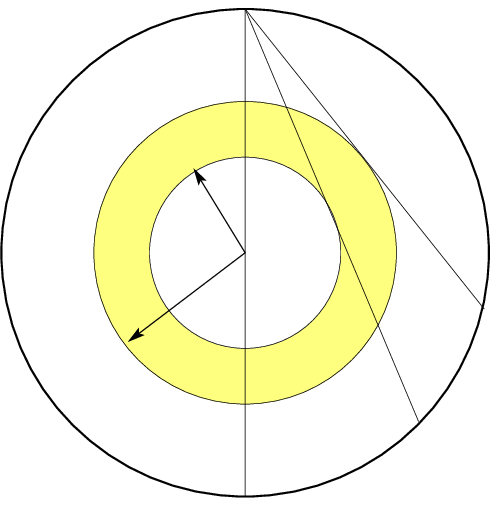}%
\end{picture}%
\setlength{\unitlength}{1184sp}%
\begingroup\makeatletter\ifx\SetFigFont\undefined%
\gdef\SetFigFont#1#2#3#4#5{%
  \reset@font\fontsize{#1}{#2pt}%
  \fontfamily{#3}\fontseries{#4}\fontshape{#5}%
  \selectfont}%
\fi\endgroup%
\begin{picture}(7898,8278)(878,-8317)
\put(6601,-5836){\makebox(0,0)[lb]{\smash{{\SetFigFont{6}{7.2}{\familydefault}{\mddefault}{\updefault}{\color[rgb]{0,0,0}$\mu_{\rm in}$}%
}}}}
\put(8776,-3511){\makebox(0,0)[lb]{\smash{{\SetFigFont{6}{7.2}{\familydefault}{\mddefault}{\updefault}{\color[rgb]{0,0,0}$\mu < \mu_{\rm out}$}%
}}}}
\put(8026,-6436){\makebox(0,0)[lb]{\smash{{\SetFigFont{6}{7.2}{\familydefault}{\mddefault}{\updefault}{\color[rgb]{0,0,0}$\mu_{\rm out} < \mu < \mu_{\rm in}$}%
}}}}
\put(6676,-2386){\makebox(0,0)[lb]{\smash{{\SetFigFont{6}{7.2}{\familydefault}{\mddefault}{\updefault}{\color[rgb]{0,0,0}$\mu_{\rm out}$}%
}}}}
\put(3751,-4936){\makebox(0,0)[lb]{\smash{{\SetFigFont{6}{7.2}{\familydefault}{\mddefault}{\updefault}{\color[rgb]{0,0,0}$r_{\rm out}$}%
}}}}
\put(5401,-8236){\makebox(0,0)[lb]{\smash{{\SetFigFont{6}{7.2}{\familydefault}{\mddefault}{\updefault}{\color[rgb]{0,0,0}$\mu > \mu_{\rm in}$}%
}}}}
\put(3676,-3286){\makebox(0,0)[lb]{\smash{{\SetFigFont{6}{7.2}{\familydefault}{\mddefault}{\updefault}{\color[rgb]{0,0,0}$r_{\rm in}$}%
}}}}
\end{picture}%

\caption{
Geometry for the uniform shell model.
The inner and outer shell radii
are $r_{\rm in}$ and $r_{\rm out}$,
respectively.
The lines of sight tangent to the inner and
outer edges of the shell are shown,
and are at nadir angle cosines
$\mu_{\rm in}$ and $\mu_{\rm out}$, respectively.
These lines of sight divide the terrestrial hemisphere
into three distinct regions,
which are labeled.
\label{fig:1shellgeom}
}
\end{center}
\end{figure}

There are three cases to be distinguished, as one can see in Figure \ref{fig:1shellgeom}:
\beq
g(\mu) =
  \left\{
  \begin{array}{ll}
    \dens_{i,0} \left( \sqrt{\mu^2-\mu_{\rm out}^2}  - \sqrt{\mu^2-\mu_{\rm in}^2}  \right) \ ,
      & \mu > \mu_{\rm in} \\
    \dens_{i,0} \sqrt{\mu^2-\mu_{\rm out}^2}  \ ,
      & \mu_{\rm out} \le \mu \le \mu_{\rm in}  \\
    0  \ , & \mu < \mu_{\rm out}
  \end{array}
  \right.
\eeq
where
\beq
\mu_{\rm in} =  \sqrt{1-x_{\rm in}^2} \ ,  \
\mu_{\rm out} =  \sqrt{1-x_{\rm out}^2} \ .  \
\eeq
These are integrable, and give
\beq
\label{eq:shellflux}
H(\mu) =
  \left\{
  \begin{array}{ll}
    \frac{\dens_{i,0}}{2} \left[ \mu_{\rm out}^2 \, h(\mu/\mu_{\rm out})
         - \mu_{\rm in}^2 \, h(\mu/\mu_{\rm in}) \right] \ ,
      & \mu > \mu_{\rm in}\\
    \frac{\dens_{i,0}}{2} \mu_{\rm out}^2 \, h(\mu/\mu_{\rm out})  \ ,
      & \mu_{\rm out} \le \mu \le \mu_{\rm in}  \\
        0
      & \mu < \mu_{\rm out}
  \end{array}
  \right.
\eeq
where
\beq
h(u) = \frac{1}{2} u \sqrt{u^2-1}
  - \frac{1}{2} \ln\left( u + \sqrt{u^2-1}  \right) \ \ .
\eeq
Note that eq.\ (\ref{eq:shellflux}) calls $h(u)$ only in the domain
$u \ge 1$, and that $h(1) = 0$.
The total flux for a uniform shell
was first calculated in \cite{kgs},
and one can easily show that
$F(\mu<1)$ reproduces their result.

There are four cases we want to illustrate;
these are plotted in Figure \ref{fig:demos}.
In the ``uniform Earth'' model, the density is the same throughout
the whole planet ($x_{\rm in} = 0$, $x_{\rm out} = 1$).
This gives an angular distribution
which grows linearly with $\mu$,
$g(\mu) = \dens_{i,0} \mu$,
and hence
$I(\theta) \propto \cos(\theta)$.
The integrated flux which increases quadratically as
$H(\mu) =  \dens_{i,0} \mu^2/2$, i.e.,
$H(\theta) \propto \cos^2 \theta$.
This therefore gives a very centrally bright
distribution.
In the ``uniform core'' model, the density is non-zero only in
a central region which extends from
$x_{\rm in}=0$ to $x_{\rm out}$.
As seen in Figure \ref{fig:demos},
this gives an inner intensity distribution which is similar
to the uniform Earth model, as one would expect,
but which goes to zero, as it should, at
the outer tangent $\mu_{\rm out}$.

As we will see below, it turns out that a more physically motivated
case is the ``uniform crust'' model, where only the
layers of the Earth (from $x_{\rm in}$ to $x_{\rm out}=1$) contribute.
This yields an intensity distribution
\beqar
g(\mu) & = &
  \left\{
    \begin{array}{ll}
    \dens_{i,0} \left[ \mu^  - \sqrt{\mu^2-\mu_{\rm in}^2}  \right] \ ,
      & \mu > \mu_{\rm in} \\
    \dens_{i,0} \mu  \ ,
      & \mu \le \mu_{\rm in}
    \end{array}
  \right. \\
H(\mu) & = &
  \left\{
    \begin{array}{ll}
    \frac{\dens_{i,0}}{2} \left[ \mu_{\rm out}^2 \, h(\mu/\mu_{\rm out})
         - \mu_{\rm in}^2 \, h(\mu/\mu_{\rm in}) \right] \ ,
      & \mu > \mu_{\rm in}\\
    \frac{\dens_{i,0}}{2} \mu^2  \ ,
      & \mu \le \mu_{\rm in}  
    \end{array}
  \right.
\eeqar
which is also linear in $\mu$, and thus
scales as $I \propto \cos(\theta)$,
for large angles.
The intensity peaks at inner tangent point $\mu_{\rm in}$,
where the line of sight is longest.
A measurement of this peak
would thus give the position of the inner edge.
Interior to $\mu_{\rm in}$,
the intensity drops to a minimum at the nadir, where
the column density is the smallest.  But the central intensity
is nonzero, and is simply related to the
peak intensity via
\beq
\frac{I_{\rm peak}}{I_{\rm center}}
 = \frac{I(\mu_{\rm in})}{I(1)}
 = \frac{\Delta s(\mu_{\rm in})}{\Delta r_{\rm shell}}
 = \sqrt{ \frac{1+x_{\rm in}}{1-x_{\rm in}} }
\eeq
This useful relation allows one to use the peripheral flux, due
to emission from Earth's outer layers,
to put a lower limit on the contribution of these
layers to the central flux.
Any observed central flux in excess of this limit
must be due to emission from the inner Earth.

The last model illustrated in Figure \ref{fig:demos}
is a single shell with arbitrary
inner and outer radii.
The intensity distribution is an amalgam of the features
seen in the  special cases of the crust and core models.
As with the crust model, the
intensity peaks at the inner tangent point,
where the line of sight is longest.
As with the core model, the intensity goes to zero at the
outer tangent.  
Thus, a measurements of the peak would give the position of the shell's
inner edge, while a measurement of the cutoff would give the outer edge.

\begin{figure}
\epsfig{file=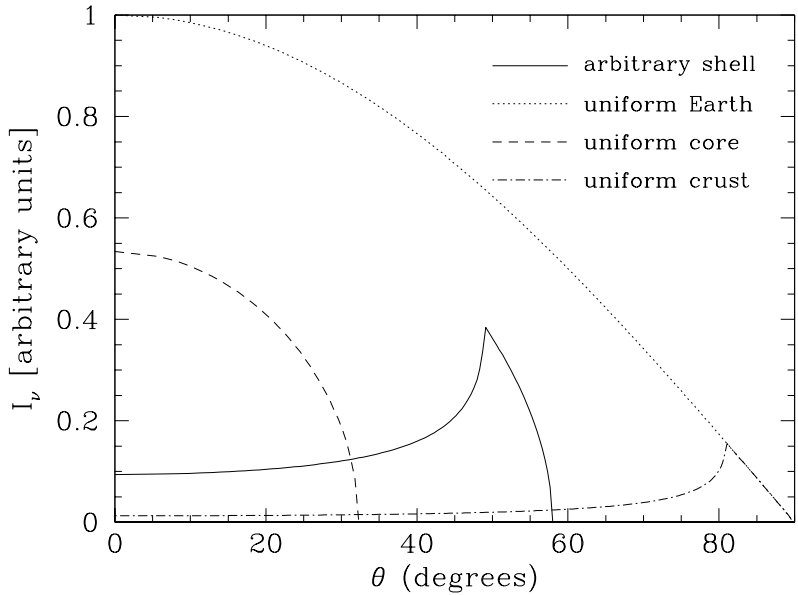,angle=270,width=0.75\textwidth}
\caption{The angular distribution of intensity
for different cases of a 1-zone, uniform density model,
shown as a function of nadir angle $\theta$.
The units in the vertical scale are arbitrary, and the
normalizations of the different curves are chosen
for clarity.
\label{fig:demos}
}
\end{figure}

\subsection{Towards a Realistic multishell Model}
\label{sect:realmod}

Based on the idealized calculations in the last section we want to
build now a model that is applicable for the Earth's Interior. The
Earth of course does not have a constant density.
However, we can approximate the density structure as a series
of uniform density shells.  Indeed, Earth models are
in practice typically expressed in this manner.
Formally, the generalization is trivial,
thanks to the lack of neutrino absorption and scattering,
which guarantees that superposition holds:
\beqar
I_i(\theta) & = &  \sum_{{\rm shells} \, j} I_{i}^{(j)}(\theta) \\
F_i(>\theta) & = &   \sum_{{\rm shells} \, j} F_{i}^{(j)}(>\theta)
\eeqar
where $I_{i}^{(j)}(\theta)$ and
$F_{i}^{(j)}(\theta)$ are the intensity and flux,
respectively, from shell $j$.

Moreover we must consider the effect of neutrino oscillations
\cite{flmrs,ntf}.
As the mass eigenstates of the neutrinos differ from the
flavor eigenstates, some of the initial \nuebar\
signal will be converted into other flavors during propagation, and thus reduce
the \nuebar\ signal in the detector.
In general, the oscillations will have a ``vacuum'' contribution,
but will be modified due to the
MSW effect, which is density- and energy-dependent.
If the MSW effect is important, then it complicates
the determination of the angular distribution,
introducing a density- and path-dependent MSW suppression factor
into eq.~(\ref{eq:master}) and its descendants.
This factor would complicate the inversion of
the angular distribution $I(\theta)$ to recover
the radioisotope source distribution $q(r)$.

However, geoneutrinos have a very low energy and as the Earth's
density is much lower compared to stellar objects we do not expect,
that the MSW effect plays any greater role. This can be justified if
the vacuum oscillation length $L_v$ is much smaller than the
oscillation length in matter $L_{\rm m}$. If we take the values of 
ref.~{\cite{msw} for the matter in the Earth and a value for 
$\Delta m^2 \sim 10^{-4} \ {\rm eV^2}$, 
we obtain $L_{\rm v} \sim 10$ km
and $L_{\rm m} \sim 1000$ km. Therefore
$|L_{\rm v}| \ll |L_{\rm m}|$, and we see that, roughly
speaking, the
vacuum oscillation length is short compared to both the
the matter oscillation length and to the changes
in density; thus the vacuum oscillations will
wash out any matter effect
and average out the pathlength dependence.
We thus follow Mantovani et al.~\cite{mcfl}
and introduce only a density-independent oscillation
probability of $(1-\frac{1}{2}\sin^22\vartheta_{12})=0.59$ in our
equations. Here $\vartheta_{12}$ is the electron-neutrino $\nu_e - \nu_x$
mixing angle; solar neutrino experiments and KamLAND
are best fit \cite{italy} 
by
$\sin^2 \vartheta_{12} = 0.315 \pm 0.035$ (and hence $\sin^2 2\theta_{12} = 0.81$).
and $\Delta m^2 = (7.3 \pm 0.8) \times 10^{-5} \ {\rm eV^2}$.
Therefore the intensity scaling of
eq.~(\ref{eq:Iscale}) changes to:
\beq
I_{i,0}^{\rm eff} =
  \left( 1-\frac{1}{2}\sin^2 2 \vartheta_{12} \right) I_{i,0}  \ \ .
\eeq
Table \ref{tab:nukeprop} sums up the important properties of the radioactive
elements in question.

In the next section we want to use this and discuss different Earth Models.

\section{Earth Models}
\label{sect:earth}

The geoneutrino intensity depends on the radioisotope density
structure $\rho_i(r)$, which is usually presented
as abundances $a_i$ and a total density $\rho$,
where $\rho_i = a_i \rho$.
The Earth's interior structure and total density
are primarily probed
via the propagation of seismic waves.
These results have been synthesized
by Dziewonski and Anderson in their Preliminary
Reference Earth Model \cite{PREM}.
Following \cite{mcfl} we will adopt this model.
In addition, for our spherically symmetric study we took
the last three kilometers of the Earth to be sediments.

The distribution of radioisotope abundances $a_i$
are obtained with different geological measurements
which, for most of the Earth's interior, are necessarily indirect.
Consequently, the abundance distribution remains model-dependent.
Indeed, the measurement of the angular distribution of geoneutrinos
(as well as the geoneutrino energy spectrum)
would provide a powerful new method to {\em measure}
the radioisotope distribution.
For the purposes of illustration here,
we will adopt the values of $a_i$ given in the reference model of
Mantovani et al. \cite{mcfl}, which are very 
detailed and draw on a wealth of geophysical data.
This model takes into account the bulk silicate
Earth model, which describes the composition of the crust-mantle
system.
Table \ref{tab:abundances}
shows the adopted abundance distribution.
Note that these are {\em elemental} abundances;
we assume that the isotopic fractions
of Table \ref{tab:nukeprop} hold throughout the Earth.
This correction is particularly important for \k40.

\begin{table}[htb]
\caption{Mantle and crust elemental abundance distribution \cite{mcfl}.}
\label{tab:abundances}
\begin{tabular}{p{3cm}|c|c|c|c}
Region  & radii [km]  & $a({\rm U})$ & $a({\rm Th})$ & $a({\rm K})$  \\
\hline
Lower Mantle & 3480--5600  & $13.2 \times 10^{-9}$ & $52 \times 10^{-9}$
  & $1.6 \times 10^{-4}$   \\
Upper Mantle & 5600--6291  & $6.5 \times 10^{-9}$ & $17.3 \times 10^{-9}$  
  & $0.78 \times 10^{-4}$  \\
Oceanic Crust & 6291--6368  & $0.1 \times 10^{-6}$ & $0.22 \times 10^{-6}$
  & $0.125 \times 10^{-2}$  \\
Lower Crust & 6291--6346.6  & $0.62 \times 10^{-6}$ & $3.7 \times 10^{-6}$  
  & $0.72 \times 10^{-2}$   \\
Middle Crust & 6346.6--6356 & $1.6 \times 10^{-6}$  & $6.1 \times 10^{-6}$  
  & $1.67 \times 10^{-2}$   \\
Upper Crust & 6356--6381 & $2.5 \times 10^{-6}$ & $9.8 \times 10^{-6}$
  & $2.57 \times 10^{-2}$   \\
Sediments  & 6368--6371  & $1.68 \times 10^{-6}$  & $6.9 \times 10^{-6}$
  & $1.7 \times 10^{-2}$    \\
Oceans  & 6368--6371  & $3.2 \times 10^{-9}$  & 0 & $4.0 \times 10^{-4}$  
\end{tabular}
\end{table}

The composition of the Earth's core deserves special attention.
The Earth's core consists largely of iron. But its density is lower
than one would expect if the core were pure iron. Therefore it is
assumed that light elements in the form of alloys are present
\cite{WMc}. So far it was generally believed, that the core does not
hold any significant amount of radioactive elements, as there was no
evidence that the radioactive isotopes in question could form alloys
with iron.
Hence, the Mantovani et al. \cite{mcfl} reference model
places radioactive elements only in the
mantle and in the crust, and are absent from the core.

On the other hand it is puzzling why solar chondrites have
a K/U ratio that is eight times higher than in the crust-mantle
system \cite{wasser}. 
In this connection, it is noteworthy that 
recent experiments demonstrate that potassium does form
alloys with iron under high temperature and pressure conditions
which likely were present at Earth's formation.
The maximum possible amounts of potassium in Earth's core
suggested by the experiments range from 60-130 ppm \cite{rama},
to 1200 ppm \cite{1200}, to as high as 7000 ppm \cite{berc}.
In light of these experiments, we will
consider the possibility of an additional radiogenic contribution from
the core, and quantify the impact of core \k40 on
the reference model.

The amount of radioactive material in the core contributes also to the
radiogenic heat of the Earth.
Presently the Earth's surface loses about 40 TW or 87 mW/$\mbox{m}^2$.
The amount of surface radiation depends on convection
and conduction properties of the Earth's interior.
The present day heat production $H$ in
units of TW of uranium, thorium and potassium with a total mass in kg
of $M_{\rm U}$, $M_{\rm Th}$, $M_{\rm K}$ respectively is
\begin{equation}
\label{eq:heat}
H = \sum_i \epsilon_i M_i
 \approx 10 \pfrac{M_{\rm U}}{10^{17} \ {\rm kg}}
   + 2.7 \pfrac{M_{\rm Th}}{10^{17} \ {\rm kg}}
   + 3.4 \times 10^{-4} \pfrac{M_{\rm K}}{10^{17} \ {\rm kg}} \ {\rm TW}
\end{equation}
where $\epsilon_i = Q_i/m_i \tau_{\beta,i}$ is the
specific non-neutrino energy loss per nucleus,
and where $M_{\rm K}$ is the {\em total} potassium mass,
for which the \k40 isotopic fraction appears in
Table \ref{tab:nukeprop}.
To obtain the heat production of potassium we used, that in 89.28$\%$
of all cases a \k40 nucleus $\beta$-decays with a average energy
of 0.598 MeV \cite{law}. The other decay mode of potassium is electron
capture with an energy of 1.505 MeV.

With knowledge of the radioisotopic content of the
Earth from geoneutrinos, eq.~(\ref{eq:heat}) can be
compared with the global heat flux, and used to determine
the Urey ratio (eq.~\ref{eq:urey}).
While the radioisotope abundances remain uncertain,
one can rather turn the problem around, and use the
global heat flux with eq.~(\ref{eq:heat}) to
set an upper limit to the radioisotopic content
and thus to the geoneutrino flux.
Such an analysis has been carried out by
Mantovani et al., who find that this ``maximal radiogenic''
model leads to fluxes about twice the level of their
reference model.

In the next section we will discuss the plots we created for the
different Earth models.

\section{Results}
\label{sect:results}

We now combine our general formalism with
various Earth models to arrive at predictions for
the geoneutrino angular distribution.
We first consider the reference model,
then its variants and its uncertainties,
and finally we comment on the effect of anisotropies
in the radioisotope distributions.

\subsection{The Reference Model}

The reference model of Mantovani et al.~\cite{mcfl}
serves as our standard and fiducial case.
The geoneutrino intensity distribution
for this model appears in Figure~\ref{fig:std-UThK}.
We see that the total intensity is peaked
near the horizon, at large nadir angles.
The  strikingly
``peripheral'' character of this
neutrino distribution
is a direct consequence of the
location of the radioisotopes in the mantle
and crust.

The experimental ability to detect this pattern
is perhaps best quantified in Figure~\ref{fig:flux},
which displays the cumulative angle-integrated flux $F$
for the reference model (c.f.~eq.~\ref{eq:flux}).
The change in the normalized flux $F/F_{\rm tot}$ over
any angle interval gives
the contribution of that interval to the total flux.
Figure~\ref{fig:flux} thus shows that fully 2/3 of the total flux
arrives in the outermost nadir angles $\theta \ga 60^\circ$.

\begin{figure}
\epsfig{file=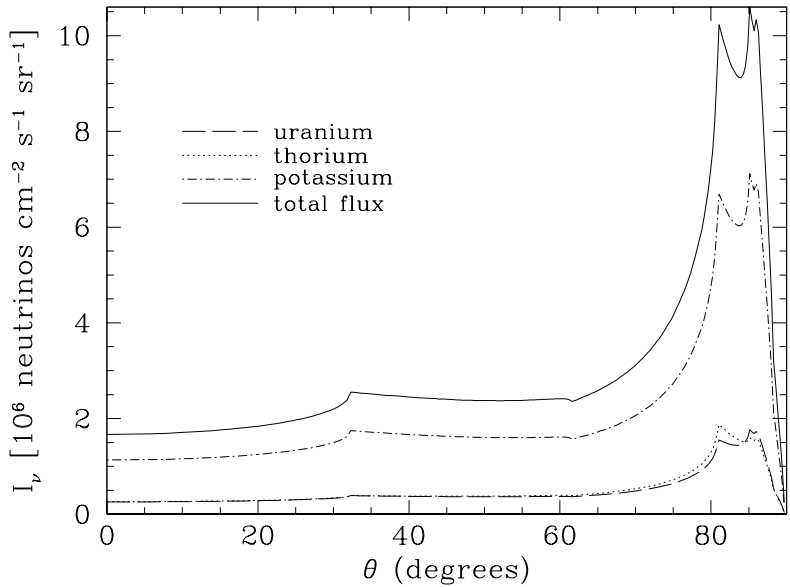,angle=270,width=0.75\textwidth}
\caption{The figure shows the reference model of ref.~\cite{mcfl}.
The dashed-dotted line represents the total expected intensity.
No contribution coming from the core has been added yet.
The other curves show the intensities from thorium, uranium and
potassium separately. It can be seen that the major contribution
is coming from potassium.
\label{fig:std-UThK}}
\end{figure}
 
This model can thus be easily tested
with an experiment having even the poorest
angular resolution.
For example, an experiment with $30^\circ$ resolution
could test whether the counts in the outer $\theta > 60^\circ$
are a factor $\sim 2$ higher than the counts in the
inner $\theta < 60^\circ$.
The confirmation of this model would vindicate
the idea that U, Th, and K are indeed lithophilic
elements that congregate in the mantle and crust.

Figure~\ref{fig:std-UThK} also shows
the expected angular distribution separately for uranium, thorium,
potassium and the cumulative intensity for the reference model.
The uranium to
thorium ratio stays approximately the same, whereas the amount of
potassium increases steeply in the crust. The intensity shows a double
peak, which is due to the fact, that the abundance of radioactive
elements is lower in the middle crust than in the upper and lower
crust.  

We note that the reference model
gives a geoneutrino distribution qualitatively
similar to the uniform crust model
presented in \S \ref{sect:formal}.
This of course traces to the positioning
of the radioisotopes in the outer Earth.
Of course the true geoneutrino behavior is more complicated,
but it does appear that for a crude analysis
(or in the presence of crude data)
the uniform crust model would be a useful
analytic approximation.

\begin{figure}
\epsfig{file=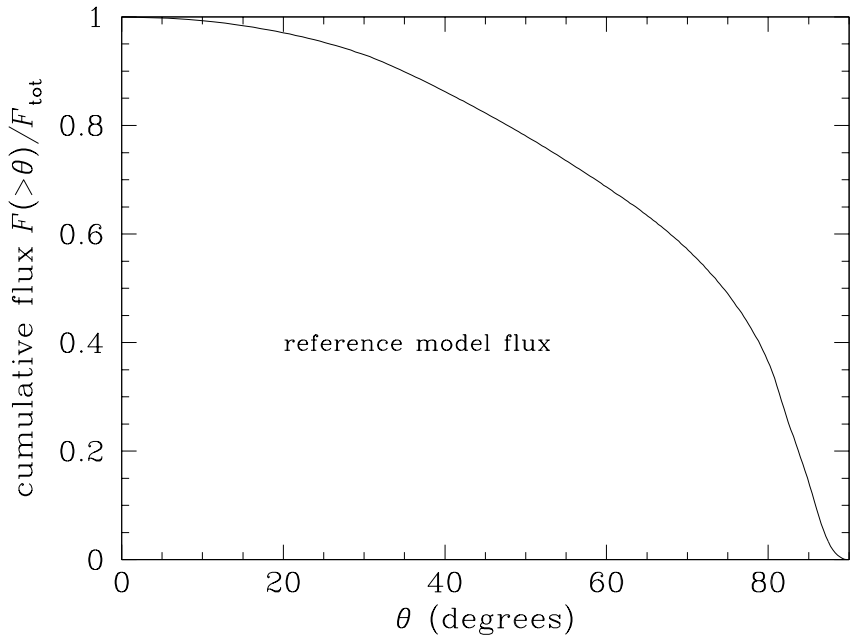,width=0.6\textwidth,angle=270}
\caption{Cumulative flux expected for the reference model.
The flux is normalized to the total,
and thus is dimensionless, spanning the range 0 to 1.
\label{fig:flux}
}
\end{figure}

Finally, we wish to comment on the effects of sediments.
We find that if the amount of sediments is increased by a
factor of 10 the overall shape of the spectrum does not change except
for a sharp peak towards $90^\circ$. An increase by a factor of two
only increases the second peak slightly. Therefore we conclude, that the
location of the detector will not change its ability to detect the
structure of the mantle and crust.

\subsection{Models with Core Potassium}

We now turn to the possibility that the Earth's core might contain
significant amounts of \k40.  Based on the experimental evidence that
potassium can form alloys with iron we use the fiducial values
obtained by different authors and add them to the reference model. We
are aware that if the abundance ratios and mass ratios of the
reference models are correct and only the total mass estimates of U,
Th and K are incorrect, then the intensity of the neutrinos coming from the
crust-mantle system should go down, but the overall shape should stay
the same. This decrease in the peripheral intensity will make the
core contribution even more dominant.
On the other hand, it is possible that
the assumptions of the reference model are
correct (incorporating the bulk
silicate Earth model with a heat production of $\sim$ 20 TW),
but the Urey ratio is closer to 1 with a large amount of
potassium in the core.
In this case it is legitimate to add the core
contribution to the reference model. In case of a fully radiogenic
model with 20 TW coming from the crust-mantle system one would expect
a total of $\sim$ 330 ppm of potassium in the core.
Figure~\ref{fig:cores} shows the angular distribution
for different amounts of potassium
in the core as could be found in the literature,
while Figure~\ref{fig:core-cumul} shows the cumulative flux
for the same models.

From Figures \ref{fig:cores} and \ref{fig:core-cumul} it is clear
that the introduction of \k40 in the core
can significantly alter the geoneutrino angular distribution.
The effect is to enhance the central intensity ($\theta \la 30^\circ$),
possibly also raising the overall detected flux.
The departure from the reference model depends of course on the
core potassium abundance.
The value given in \cite{berc} of 7000 ppm potassium in the core is
only an upper limit, but in this case the core would clearly dominate
the distribution. For 1200 ppm of potassium in the core \cite{1200}
the maximum intensities coming from crust and core are approximately
the same. But even the much lower value obtained in \cite{rama} of
60-130 ppm might
still be detectable with a future neutrino detector.

A future low energy neutrino detector with angular
resolution will be able to distinguish between the opposing
models.
Again, with only a modest resolution, it will
already be possible to make important statements.
An experiment with $30^\circ$ resolution
could divide the emission into central, medial, and peripheral
bins, and the relative counts would test both the
concentration of radioisotopes in the mantle and crust, as well
as the possible presence of \k40 in the core.

\begin{figure}
\epsfig{file=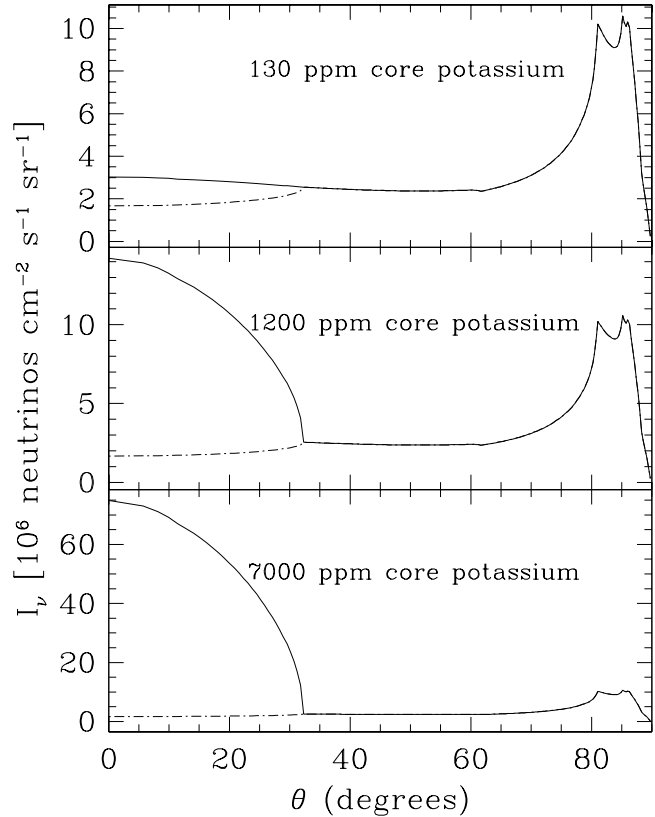,width=0.6\textwidth}
\caption{
The curves show the possible abundances for potassium in the core as
found in \cite{rama}, \cite{berc}, \cite{1200}. The dashed-dotted line
is the intensity of the reference model.
\label{fig:cores}
}
\end{figure}

\begin{figure}
\epsfig{file=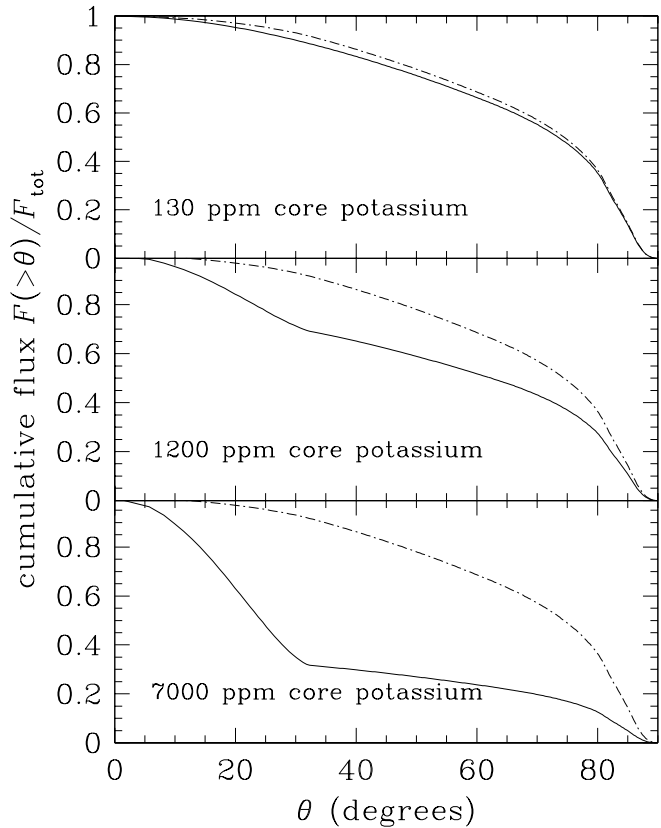,width=0.6\textwidth}
\caption{
The plots are similar to Fig.~\ref{fig:cores},
except that they show the cumulative
flux, normalized as in Fig.~\ref{fig:flux}.
The dashed line is the flux of the reference model.
\label{fig:core-cumul}
}
\end{figure}

\subsection{Uncertainties}

We want to investigate the impact of uncertainties of the reference model
on the geoneutrino distribution.
The uncertainties for crust and mantle are
independent of each other.
The crust itself can vary by about a factor of 2
in radioisotope abundances and thus in geoneutrino intensity
\cite{mcfl}.
The net effect of these variations thus depends on
how the crust and mantle uncertainties combine.

\begin{figure}
\epsfig{file=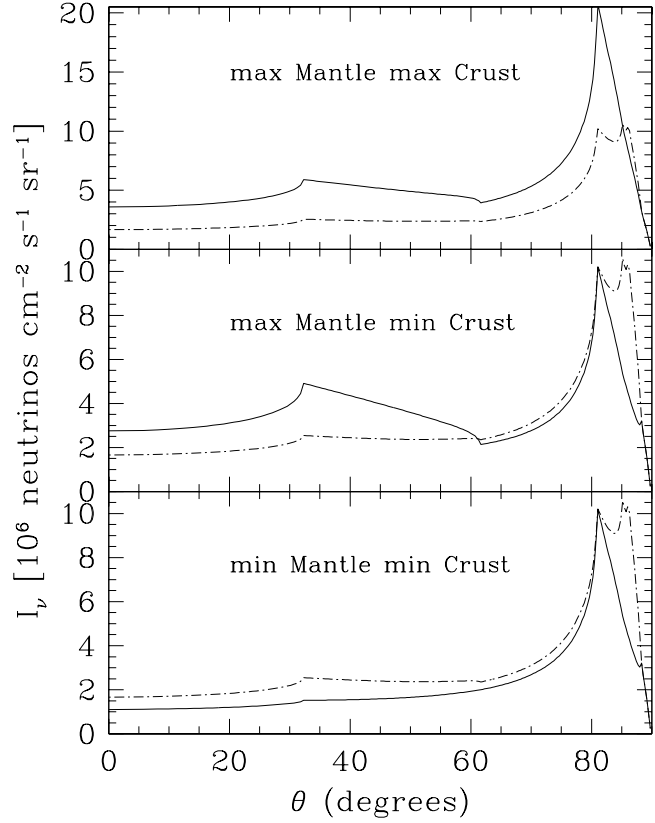,width=0.6\textwidth}
\caption{
The upper plot shows the intensity with a maximal amount of radioactive elements in crust and mantle. In the lower panel the
intensity for a minimal abundance in crust and mantle is plotted. As
the uncertainties in the abundances of mantle and crust are
independent of each other \cite{mcfl}, the plot in the middle shows a
hybrid scenario with a maximal abundance in the mantle, whereas the
crust abundance is minimized. The dashed-dotted line in all three
plots is the reference model intensity is added as dashed dotted line
in all three plots as comparison. It can be seen that the reference
model is rather at the low side of the possible range of
abundances. That the crust is only represented by a single peak
is an artifact which arises because the
uncertainties given in \cite{mcfl} do not distinguish between the
different layers of the crust;
this changes the intensity in the
outer crust significantly.
\label{fig:uncertainty}
}
\end{figure}

Figure~\ref{fig:uncertainty} shows some of the
possible uncertainties.
In the plot showing the minimum amount of
radioactive elements in the crust-mantle system it can be perceived, that the
reference model is on the low side of the possible abundances in
comparison to the maximum abundances, where the intensity grows by a
factor of two. The absence of a double peak in the intensity is due to
the fact, that in \cite{mcfl} only a uncertainty for the whole crust
is given, which does not take into account the distinction between
lower, middle and upper crust. The overall shape of the maximum and
minimum abundance plots stays overall very similar to the reference
model, although there is a deviation from the general form at angles
$\theta \ga 30^\circ$.
Thus we see that in all cases there appears a
large peripheral flux, which remains a robust
and highly testable prediction of this model.

For a more detailed analysis it will be necessary to construct a model
of the outermost layer of the Earth, as we assumed for a change in
altitude an increase in the number of neutrinos coming from the
sediments. But in any case the crudeness of our estimate
only infects the results for the outer periphery, and
the central angles remain reliable, as we now see.

\subsection{Crust Anisotropies and Observing Strategies}

We have assumed spherical symmetry throughout,
and thus our calculations cannot directly address
the effect of anisotropies in the radioisotope distributions.
Yet these anisotropies, which reside in the crust,
can have a very significant impact \cite{mcfl},
since the crust is the largest radioisotope reservoir.
For example, Mantovani et al.~\cite{mcfl}
predict fluxes which differ by more than a factor of
2 between locations above minimum and maximum crust
depths.
This would lead to changes in the overall geoneutrino
intensity, and possibly to observable azimuthal
asymmetry.

We note that emission from the crust
will affect the intensity only at the largest nadir angles.
Thus we expect our spherical calculation
to be reliable at small to median angles.
Furthermore, one can get a rough understanding
of the effect of different crust depths by
adopting spherical models in which the crust layers
have the properties appropriate for the experimental location.

To consider an extreme example,
consider a hypothetical model where we
assume that the neutrino detector is on a ship. That means that the
crust has the abundances of the oceanic crust and the top layer is
three kilometers of ocean water. Of course in this case there is no
sufficient shielding from cosmic rays and atmospheric radiation, which
will make the detection of low energy neutrinos hard if not
impossible. But never the less the effects are interesting, as the
contribution coming from large angles is reduced by a factor of three,
which should make it easier to detect the effects coming from the core
and mantle. We have presently no model for an  anisotropic Earth, but we
expect, that for the real Earth in the described case the intensity
for lower angles ($\theta \la 60^\circ $) will be very similar to the
reference model (in case of its correctness), only the contribution from the angles $\ga 30^\circ$
will be noticeably reduced.

\section{Discussion}
\label{sect:conclude}

Geoneutrinos offer invaluable and unique information
about the energetics and structure of the Earth
\cite{kgs,fmr}.
The longstanding dream of measuring this neutrino
population has now begun to be realized
with the first detection of a geophysical
signal by the KamLAND experiment \cite{kamland}.
This achievement is already a triumph, as
the geophysical component is (by design!) dominated by
the signal from reactor neutrinos.
Nevertheless, we believe it is now worthwhile to
look forward to the
even more challenging results of determining
the angular distribution of geoneutrinos,
which offers a wealth of new information.

In this paper we thus have calculated the angular distribution
of geoneutrinos
which arise in $\beta$-decays of potassium, thorium,
and uranium.
We have developed the general formalism for
the neutrino intensity in a spherically symmetric Earth.
We find that the geoneutrino angular distribution, once known, can
be inverted to fully recover the terrestrial radioisotope
distribution.  Thus the geoneutrio ``sky'' can provide
a tomography of the Earth's structure, and yields
the full radial dependence of the radiogenic heat production.

Turning to model-building,
we explore the idealized case of an arbitrary shell of
uniform density.
This can be generalized to give an Earth model which
is a series of uniform shells.
We then adopt the
radioisotope profile of the Mantovani et al.\ reference model
\cite{mcfl} and calculate the resulting
angular distribution
(Fig.~\ref{fig:std-UThK}).
Because the reference model places
all radioisotopes in the mantle and crust,
the resulting geoneutrino intensity is highly
``peripheral,'' with
2/3 of the flux coming from nadir angles
$\theta \ga 60^\circ$.
Thus, even a crude measurement of the
angular distribution (say, in three $30^\circ$ bins)
would strongly test this prediction.

We have also investigated the effect of physically
plausible variations to the reference model.  
Mantovani et al.\  \cite{mcfl} identify uncertainties in their
radioisotope distributions which have the effect of
multiplicatively raising or lowering the neutrino intensity,
without a significant change to the angular shape.
However, both the shape and normalization of the intensity
can change strongly if the Earth's core contains
a significant amount of potassium, contrary to the assumptions  
of the reference model.
If core potassium abundances are near the current
upper limits \cite{berc},
the resulting geoneutrino signal can dominate the
total flux.  
Measurement of the angular distribution will probe the
radioisotope abundances in the core.
In particular, measurement of the intensity
inside a nadir angle $\theta \la 30^\circ$ can
discriminate among possibilities recently suggested
in the literature.

Moreover, the central and
peripheral intensities are related, since the peripheral
neutrinos come from outer shells which also contribute to
the central signal.  Thus a measurement
of the peripheral flux can be used to place
a lower limit to the central flux.
A difference between this lower limit and the
observed central flux then amounts to a
detection of some radioisotopes in the core.

A determination of the geoneutrino angular distribution
will also solidify the connection between the geothermal
heat flux and the geoneutrino flux.
As noted by \cite{fmr}, the {\em radial}
component of the neutrino flux is directly related,
by Gauss' law, to the geothermal heat production
(eq.~\ref{eq:heat}).
But neutrino emission is locally isotropic
and hence contains non-radial components;
thus the angle-integrated geoneutrino flux
in turn includes the non-radial components,
and thus can only be related to the heat flux given
a model of the radioisotope density profile.
However, a measurement of the
geoneutrino angular distribution
can be inverted to recover the terrestrial
radioisotope density distribution.
This will allow for a full calculation not only of
the global radiogenic heat production, but also
of its radial dependence.  Thus one can
test in detail models of heat production and transport.

In this way, the neutrino intensity can be used to
measure the radioactive contribution to
the geothermal heat flux.
This can then be compared to geophysical measurements
\cite{stein} of the total heat flux.
A comparison of these results will yield a new and
robust measurement of the Urey ratio (eq.~\ref{eq:urey})
\cite{richter}.
This in turn will shed light on the
thermal history of the Earth, and quantify
the importance of any non-radioactive heating,
presumably due to residual ``primordial'' processes
during the formation of the Earth.

Also, we note that the reference model we have adopted
normalizes to the observed terrestrial heat flux
and a Urey ratio of 0.5.
Thus, if this Urey ratio is correct, but the
heat flux contains a significant component from the
core, this will reduce the contribution from the
outer layers, which will act to redistribute the
peripheral intensity to the interior.

These particular results illustrate a larger more general
conclusion, that
geoneutrinos open a new window to the Earth's interior.
Measurement of the angular distribution of
geoneutrinos will allow us to infer the
radioisotope distribution of the Earth.
This in turn offers a new
probe of the Earth's structure--for example,
allowing a test of how sharp the
radioisotope boundaries are in going from crust to mantle to core.
If the boundaries are sharp, then the angular distribution
will offer unprecedented new measures of the positions
of these boundaries and thus will be a general probe
of the interior structure of the Earth.

What are the practical prospects of measuring
the geoneutrino angular distribution?
A complete discussion is beyond the scope of this paper,
but it seems to us that there is some reason for hope.
Antineutrinos are usually detected via
inverse $\beta$-decay of protons,
$\bar{\nu}_e + p \rightarrow e^+ + n$,
and the subsequent observation of the rapid $e^+$
annihilation signal as well as the delayed $n$ capture
onto an ambient nucleus.
Part of the momentum of the neutrino is
transferred to the neutron, which thus contains
directional information.
This directionality is however lost if
the neutron scatters before it is captured.
Thus, directional sensitivity is best for
liquid scintillators which use materials (such
as gadolinium) which a large neutron capture cross section.
At present, no such measurement appears
possible: KamLAND is a liquid scintillator experiment,
but the scintillator is an organic material
which does not have a large neutron capture cross section.
However, the reactor neutrino experiment CHOOZ
did use gadolinium scintillator, and
was able to reconstruct the source direction to within
about $\sim 20^\circ$ \cite{chooz}, which as we have emphasized would
already be geophysically interesting.
While CHOOZ is no longer operating, and was
too small to detect the geoneutrino intensity, it nevertheless demonstrated
that antineutrinos with an energy spectrum similar to
that of geoneutrinos can be already detected with
modest angular sensitivity.

We thus
strongly hope
that experimental effort be undertaken
to measure the angular distribution of geoneutrinos.
In particular, we suggest the construction of a scintillator antineutrino
experiment using a medium with a large neutron capture cross section,
such as gadolinium.
Indeed, there has been recent interest in
somewhat similar experiments to measure
the higher-energy signal from
the cosmic supernova antineutrino background \cite{gadzooks,oberauer}.
While the geoneutrino flux is significantly higher than
the supernova background,
geoneutrino detection will be all the more challenging given that typical
geoneutrino energies are
relatively low, $\sim 0.5 - 1.5$ MeV \cite{kgs,mcfl}.
And of course, a good geoneutrino experiment clearly will need to
be as far as possible from reactors (and possibly submarines!).
This is particularly important if the radioisotope
distribution really is concentrated in the outer
Earth, as this will lead to a peripheral flux which
can be confused with (but possibly calibrated by) the contribution from
nearby detectors.

We note as well that we have so far considered the
total, energy-integrated, intensity.
However, neutrino energy information is also available,
and indeed geoneutrino spectra have been presented
\cite{mcfl}.
Since the emitted energy spectra take the well-understood
$\beta$-decay form,
with sufficient energy resolution it is possible to
separate the U, Th, and K components.
If this can be done in conjunction with even
the crudest angular resolution, it would
be possible to actually distinguish
beyond all doubt between neutrinos coming from U, Th and K
and to obtain a particularly complete picture of the
radioactive Earth.

Thus we believe the quest to measure an image of the ``geoneutrino sky''
is a worthy if challenging goal.
As we have shown, even the first, crudest attempts
at neutrino imaging will yield important results,
and will thus impel further improvements.
It is our hope that this paper encourages
these efforts.

\acknowledgments
We are grateful to
Stuart Freedman for very helpful discussion
regarding experimental issues and prospects regarding
antineutrino directional sensitivity.
We thank Charles Gammie for encouragement and for
alerting us to the Abel transform, and
V.~Rama Murthy for enlightening discussions
regarding core potassium.
The work of BDF
is supported by the National Science Foundation grant AST-0092939.

\appendix

\section{Terrestrial Tomography: Inverting the Angular Distribution}
\label{sect:tomography}

We may write the intensity distribution (eq.~\ref{eq:master})
in dimensionless units as
\beq
\label{eq:Isigma}
I(\sigma)
  = 2 R_\oplus \int_{\sigma}^{1} dx \ \frac{x \ q(x)}{\sqrt{x^2-\sigma^2}}
  \equiv \int_0^\sigma dx \ K(\sigma,x) \ q(x)
\eeq
where
$\sigma = \sin \theta \in [0,1]$ and
$x = r/R_\oplus \in [0,1]$.
Thus both $I$ and $q$ are defined on
the interval $[0,1]$.
Furthermore, for an experiment on the
surface of the Earth, we expect that $I(1) = 0 = q(1)$ because the
Earth's density goes to zero at the surface (by definition!).
However, a real experiment located
slightly under the surface of the
Earth might have a nonzero horizonal flux $I(1)$.

Clearly, $I(\sigma)$ is an integral transformation,
with $K(\sigma,x) = x/\sqrt{x^2-\sigma^2}$ the kernal.
Specifically, eq.~(\ref{eq:Isigma})
is a version of the Abel transform \cite{bracewell}.
In fact, the usual Abel transform is applied
to a function defined over an infinite domain,
but fortunately one can show that the key results
carry over to our case of a finite domain.

The inverse Abel transform appropriate for our case  is
\beqar
\label{eq:inverse}
q(x) & = & - \frac{1}{\pi R_\oplus}
   \int_x^1 d\sigma \frac{I^\prime(\sigma)}{\sqrt{\sigma^2-x^2}} 
   + \frac{I(1)}{\pi R_\oplus \sqrt{1 - x^2}} \\
 & = & -  \int_0^{\sqrt{1-x^2}} d\mu \frac{I^\prime(\mu)}{\sqrt{1-x^2 - \mu^2}}
   + \frac{I(\mu=0)}{\pi R_\oplus \sqrt{1 - x^2}} \\
\eeqar
where $\mu = \cos \theta$,  and
$I^\prime(y) = dI(y)/dy$ is the usual
derivative.

Equation (\ref{eq:inverse})
thus demonstrates by construction that, given a complete knowledge of the
intensity distribution, one can fully recover the
radioisotope source distribution.
Thus, measurement of the geoneutrino angular distribution
truly does carry the promise of tomographic imaging
of the Earth's interior.
In addition, with $q(x)$ in hand,
one can completely determine the radiogenic heat production
of the Earth, both globally and as a function of depth.

Furthermore, eq.~(\ref{eq:inverse})
has the properties one would expect on physical grounds.
The density at $r = R_\oplus x$ depends only
on the intensity derivative for the region $\sin \theta \ge x$,
i.e., angles along or exterior to the tangent angle.
Thus, inferring the outer density structure
requires only knowledge of the peripheral intensity.
On the other hand, to recover the inner density structure
requires both peripheral and central intensities.
This is indeed sensible if one thinks of the
angular distribution roughly as a linear combination
of intensities along the line of sight:
outer angles have only a few ``terms'' in the sum,
while inner angles contain all ``terms.''

One consequence of this result is that the peripheral
intensity constrains the central intensity, by
setting a lower limit on it.
If we consider $I(\sigma)$ only for $\sigma > \sigma_0$,
we can infer a lower limit $q_{\rm min}$ to the density
distribution at $x < \sigma_0$, namely
\beq
q(x) > q_{\rm min}(x) = - \frac{1}{\pi R_\oplus}
   \int_{\sigma_0}^1 d\sigma \frac{I^\prime(\sigma)}{\sqrt{\sigma^2-x^2}}
\eeq
This example illustrates that even with an incomplete
or low-resolution
determination of the intensity pattern,
one can draw powerful physical conclusions.

\end{document}